\documentclass{aa}  

\usepackage{graphicx}
\usepackage{txfonts}

\newcommand{\teff}{T$_{\rm eff}$}
\newcommand{\logg}{$\log g$}
\newcommand{\lsim}{\raisebox{-0.6ex}{$\stackrel{{\displaystyle<}}{\sim}$}}
\newcommand{\gsim}{\raisebox{-0.6ex}{$\stackrel{{\displaystyle>}}{\sim}$}}

\def\black{
}

\begin{document}

\title{White dwarfs in the Capodimonte deep field
\thanks{
Based on observations obtained at the following ESO instruments/telescopes:
WFI@2.2m, EFOSC2@3.6m and EMMI@NTT under proposals 63.O-0464(A), 64.O-0304(A),
65.O-0298(A), 68.D-0579(A), 69.D-0653(A).}
}

\author{R. Silvotti\inst{1}, S. Catal\'an\inst{2}, M. Cignoni\inst{3}, 
J.M. Alcal\'a\inst{1}, M. Capaccioli \inst{4,5}, A. Grado\inst{1}, 
M. Pannella\inst{6}
%
%
}

\offprints{R. Silvotti}

\institute{INAF--Osservatorio Astronomico di Capodimonte, via Moiariello 16,
80131 Napoli, Italy\\
\email{[silvotti;jmae;agrado]@na.astro.it}
\and
Institut de Ci\`encies de l'Espai, (CSIC-IEEC), Facultat  de Ci\`encies, 
Campus UAB, 08193 Bellaterra, Spain\\
\email{catalan@ieec.uab.es}
\and
Dipartimento di Astronomia, Universit\`a di Bologna, via Ranzani 1, 
40127 Bologna, Italy\\
\email{michele.cignoni@unibo.it}
\and
Dipartimento di Scienze Fisiche, Universit\`a Federico II, via Cintia, 
80126 Napoli, Italy\\
\email{capaccioli@na.infn.it}
\and
INAF--VSTceN, via Moiariello 16, 80131 Napoli, Italy
\and
National Radio Astronomy Observatory (NRAO), 
P.O. Box O, 1003 Lopezville Road, Socorro, 
NM 87801-0387, U.S.A.\\
\email{mpannell@nrao.edu}
}

\date{Received 26 November 2007 / Accepted 21 November 2008}

\abstract
{}
{In this article we describe the search for white dwarfs (WDs) in the
multi-band photometric data of the Capodimonte deep field survey.}
{The WD candidates were selected through the $V-R_C$ vs $B-V$
color-color diagram. For two bright objects, the WD nature has been confirmed
spectroscopically, and the atmospheric parameters (\teff~ and \logg) 
have been determined. 
We have computed synthetic stellar population models for the
observed field and the expected number of white dwarfs agrees with the 
observations. 
The possible contamination by turn-off and horizontal branch
halo stars has been estimated.
The quasar (QSO) contamination has been determined by comparing the number of 
WD candidates in different color bins with state-of-the-art models and previous
observations.}
{The WD space density is measured at different distances from the Sun.
The total contamination (non-degenerate stars + QSOs) in our sample is 
estimated to be around 30\%.
This work should be considered a small experiment in view of more ambitious 
projects to be performed in the coming years in larger survey contexts.}
{}


\authorrunning{R. Silvotti et al.}
\titlerunning{White dwarfs in the Capodimonte Deep Field}

\maketitle


\section{Introduction}

The intrinsic faintness of white dwarfs (WDs) means that they are only seen
at small distances from the Sun and that their statistical properties are 
still not well known.
A complete sample of white dwarfs only exists in a small volume with a radius
of 13~pc.
Based on this sample of 43 stars, and adding another 79 objects with 
known distances all within 20~pc from the Sun, Holberg et al. 
(\cite{holberg08}) obtained a WD local space density of 
$(4.8\pm0.5)\times10^{-3}$~pc$^{-3}$.

Statistical studies of white dwarfs have been increasing in the past years,
thanks to the results of recent and/or ongoing surveys, in particular the Sloan
Digital Sky Survey (SDSS, Eisenstein et al. \cite{eisenstein06}), 
which has roughly doubled the number of spectroscopically confirmed white 
dwarfs, with about 6,000 new discoveries, allowing detailed studies of the 
mass distribution (Kepler et al. 2007) and the mass and luminosity function 
(De~Gennaro et al. 2008 and references therein).
However, the completeness limit of the SDSS photometric data, around
g$\approx$19.5 (De~Gennaro et al. 2008), is not deep enough 
to study the WD distribution of these stars across the galactic disk, 
and their scale height, in particular for the coolest objects.
Deeper samples exist, for example the WD candidates identified in the 
Canada-France-Hawaii Telescope Legacy Survey (CFHTLS, Limboz et al. 2007), 
even though they are generally limited to small areas 
(3.6 square degrees for the CFHTLS WDs).
Moreover, color selection alone is not enough to identify cool white dwarfs.
For white dwarfs cooler than about 8,000~K, astrometry is essential 
when spectroscopic data are not available and the reduced proper motion diagram
(Luyten 1918) is the best way to separate these objects from metal-weak, 
high-velocity, 
main-sequence Population II subdwarfs (Kilic et al. 2006).

Of particular interest are the so-called ultra-cool white dwarfs 
(\teff\lsim4,000~K), very old objects 
that contain precious information on the primordial epoch of our galaxy.
These objects fall near the deep minimum of the WD luminosity function,
at L/L$_{\odot} \simeq -4.5$, caused by the finite age of the galactic disk.
Measuring the position of this minimum, we can measure the age of the galactic 
disk itself (D'Antona \& Mazzitelli 1978, 
Harris et al. 2006 and references therein).
Presently, the number of known ultra-cool WDs is about 20 (Harris et al. 2008)
and most of them were discovered in the Sloan Digital Sky Survey.

For the future, one of the most ambitious projects that study the WD 
populations is the ESA Gaia astrometric space mission, 
which will be able to discover about 400,000 white dwarfs with a 
detection rate close to 100\% up to 100 pc (Jordan 2007).

In this article we describe the search for white dwarfs in the
multi-color photometric data of the Capodimonte deep field survey.
In sect. 2 we present the selection criteria of our sample, based on 
morphological classification of point sources and colors.
In sect. 3 we describe the spectroscopic results on three bright targets.
In sect. 4 synthetic stellar population models are computed and the expected
number of white dwarfs is compared with our sample.
Contamination from turn-off and horizontal branch halo stars is estimated 
and discussed.
QSO contamination is considered in sect. 5, where our results
are discussed and compared with previous findings.


\section{Photometry: selecting WD candidates}

The ``Osservatorio Astronomico di Capodimonte'' Deep Field 
(OACDF, Alcal\'a et al. 2004) is a multi-band ($B,V,R_C$\, plus shallow $I_C$) 
photometric survey covering 0.5 square degrees at high galactic latitude
(RA$_{2000}$$\simeq$12:25, DEC$_{2000}$$\simeq$--12:49 or 
$l\simeq$293.0, $b\simeq$49.56 in galactic coordinates),
performed using the Wide Field Imager (WFI) attached to the ESO 2.2~m telescope
at La Silla observatory.
The 5~$\sigma$ limiting magnitudes are: 
B$_{AB}$=25.3, V$_{AB}$=24.8 and R$_{AB}$=25.1.
Typical errors (including source extraction, zero-points and extinction terms)
at magnitude 20 (23) are 0.06 (0.10) mag in the $B$ band and 
0.05 (0.08) mag in $V$ and $R_C$.

\subsection{Extra-galactic extended contaminants}

From the original catalogs of the OACDF survey, we optimized the selection 
criteria to isolate point source objects from extended ones, using the 
{\it flux radius} parameter from SExtractor (Bertin \& Arnouts 1996).
{\it Flux radius} is the aperture radius in pixels 
(1~pix~=~0.238~arcsec for WFI) where a defined fraction of the light
(50\% in our case) is collected. 

Fig.~1 represents the {\it flux radius} versus magnitude, where the stellar
branch can be easily identified.
The six panels refer to the $BVR_C$ bands in two adjacent WFI fields (OACDF2 
and OACDF4).
We can see that the contamination from extended objects starts typically near 
magnitude 21-22, depending on the photometric band considered, and becomes 
quite strong at magnitude $\sim$23 (or $\sim$24 in the $B$ band). 
The selected objects (black points) are those falling inside the boxes indicated
in Fig.~1. This criterion must be valid for the three $BVR_C$ photometric bands
simultaneously.
In this way we can exclude most of the extra-galactic extended sources 
and saturated stars that lie in the left upper part of each box.

\begin{figure}[t]  
\vspace{8.7cm}
\includegraphics{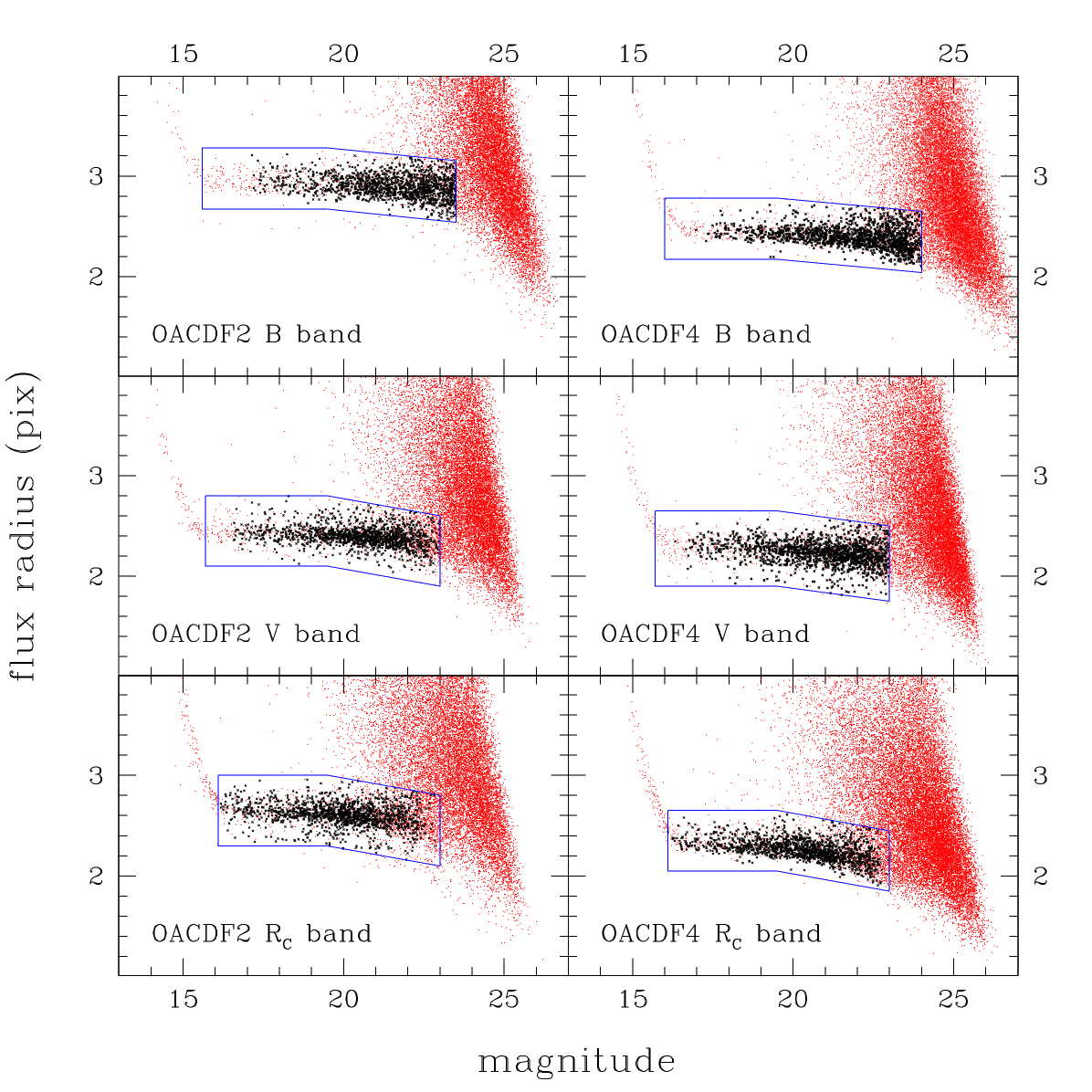}
\caption[]{Selection of point sources objects from the OACDF catalogs.
The polygons represent the regions where the point-like objects are concentrated
(see the text for more details).}
\label{flux}
\end{figure}

\subsection{WD candidates from the color-color diagram}

\begin{figure}[h]  
\vspace{8.7cm}
\includegraphics{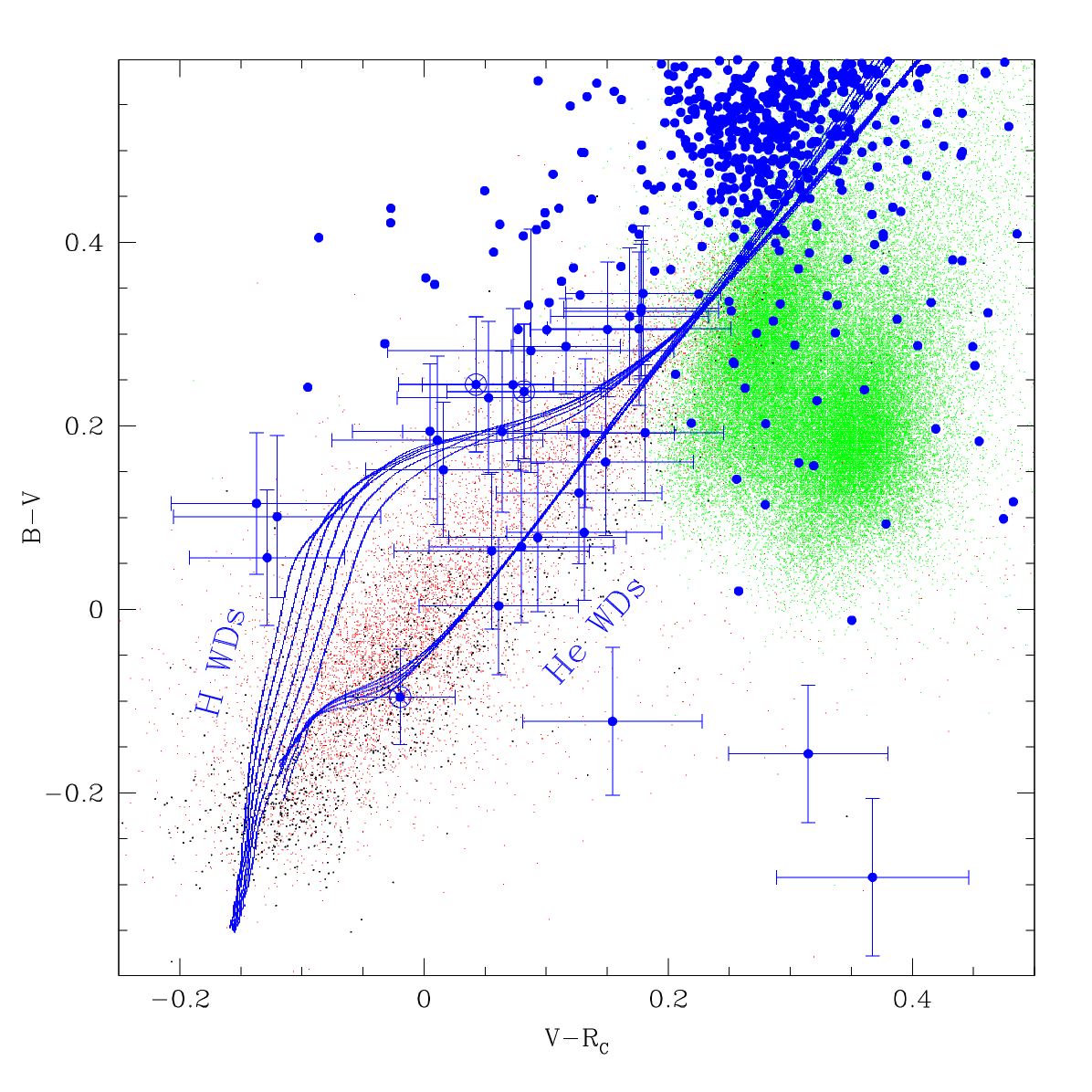}
\caption{Color-color plane of the OACDF point source objects (blue dots)
compared with Bergeron's theoretical WD tracks with
$\log g=7.0,7.5,8.0,8.5,9.0,9.5$ (H WDs) or $\log g=7.0,7.5,8.0,8.5,9.0$
(He WDs) from {\it http://www.astro.umontreal.ca/~bergeron/CoolingModels/} 
(see Holberg \& Bergeron 2006 and references therein for more details).
The objects with error bars represent the WD candidates with $B-V<0.35$.
The three circled points represent
the WD candidates for which spectroscopy was performed (see Section 3).
Since galactic reddening is rather small (at most E$(B-V)=0.05$  and 
E$(V-R_C)=0.03$ mag for the hotter objects, more distant from us and outside
the disk), extinction is not considered in this plot.
The green, red and black points represent, respectively, all the SDSS quasars, 
white dwarfs and hot subdwarfs (see the text for more details).
Note that the black symbols of the hot subdwarfs are slightly larger in order
to enhance these rare objects.
On the top of the QSO cloud, near the upper right corner, the clump of
OACDF red dwarfs is visible.
}
\label{colors}
\end{figure}

To select the white dwarf candidates, in Fig.~\ref{colors} the 
position of the point-like sources in the ($V-R_C$, $B-V$) plane is compared 
with the theoretical WD tracks from Holberg \& Bergeron (2006) 
and with the location of
the SDSS quasars (Data Release 5, Schneider et al. 2007), white dwarfs 
and hot subdwarfs (Data Release 4, Eisenstein et al. 2006).
The Sloan colors of the SDSS objects were converted to the 
Johnson-Cousin system using the color transformations
of Jester et al. (2005).
The small discrepancy between WD tracks and hot SDSS white dwarfs (the latter
being $\sim$0.05 redder in $V-R_C$) is due to the fact that Jester's equations
are not optimized for very blue objects. We have verified that in a
($g-r$, $r-i$) plane, the agreement between WD tracks and hot SDSS white dwarfs
is much better (Bergeron's evolutionary tracks are available also in Sloan 
colors).

Considering only the hottest objects ($B-V<0.25$), there are 21 WD 
candidates whose location is compatible with H or He white dwarfs 
having an effective temperature higher than about 8,400 or 8,100~K
respectively.
Another $\sim$19 objects falling on the right side of the WD cooling tracks 
could be white dwarfs with a red companion.
However, in this region contamination from quasi-stellar objects (QSOs) is 
very strong at $B-V$~\gsim~$0$ (see Fig.2),
and therefore most of these objects must be QSOs.
For three of them, having very blue colors ($B-V<-0.1$), we can exclude a QSO
nature: these objects can be either white dwarfs or hot subdwarfs with 
a cool companion. We know that a significant fraction 
of hot subdwarfs ($\sim$50\%
or even more for subdwarf B stars, Morales Rueda et al. 2006) reside 
in binaries; however hot subdwarfs are much more rare than white dwarfs
(see Fig.2 and caption). 
Therefore these 3 objects are included in our list of WD candidates.
Extending our color limit to $B-V=0.35$ (corresponding to 
\teff $\simeq$ 7,150~K
for both H and He WDs), the total number of WD candidates 
increases by a factor of $\sim$2 but the selection becomes more 
difficult because in this region the WD cooling tracks are very close
(almost inside) the QSO clump.
At $B-V$\,\gsim\,$0.25$, also the contamination from OB stars
(main sequence objects and horizontal branch stars) starts.
As a general criterion we have considered as WD candidates the objects 
whose error bars intersect the WD cooling tracks, excluding the regions where
the QSO density dominates.
In this way we add 8 more WD candidates, bringing to 32 the total number 
of WD candidates with $B-V<0.35$.

At $B-V$\,\gsim\,$0.35$\,-\,$0.40$, stellar contamination becomes important,
in particular from halo objects (see section 4).
In order to be identified, cool white dwarfs require astrometric 
data and the use of the reduced proper motion diagram.
These objects are not considered in this article.


\section{Spectroscopy of three bright WD candidates}

Among the WD candidates, three bright stars, marked with a circle in Fig.~2,
were selected to perform spectroscopic follow-up.
The spectroscopic observations were carried out at La Silla observatory
using EMMI at the NTT and EFOSC2 at the 3.6~m telescope, 
both with the MOS (Multi-object spectroscopy) configuration.
In the case of EMMI, the grism number 5 was used, with a spectral coverage 
from 4000 to 6600 \AA, and a resolution of 5~\AA~ FWHM (implying a resolving
power R$\sim$1100 at central $\lambda$). 
In the case of the EFOSC2 instrument, the grism number 1 was used, covering the
spectral range from 3200 to 9000 \AA~ with a resolution of 40~\AA~ FWHM 
(R$\sim$150 at central~$\lambda$). 

\begin{figure}[h,t]  
\vspace{12.5cm}
\includegraphics{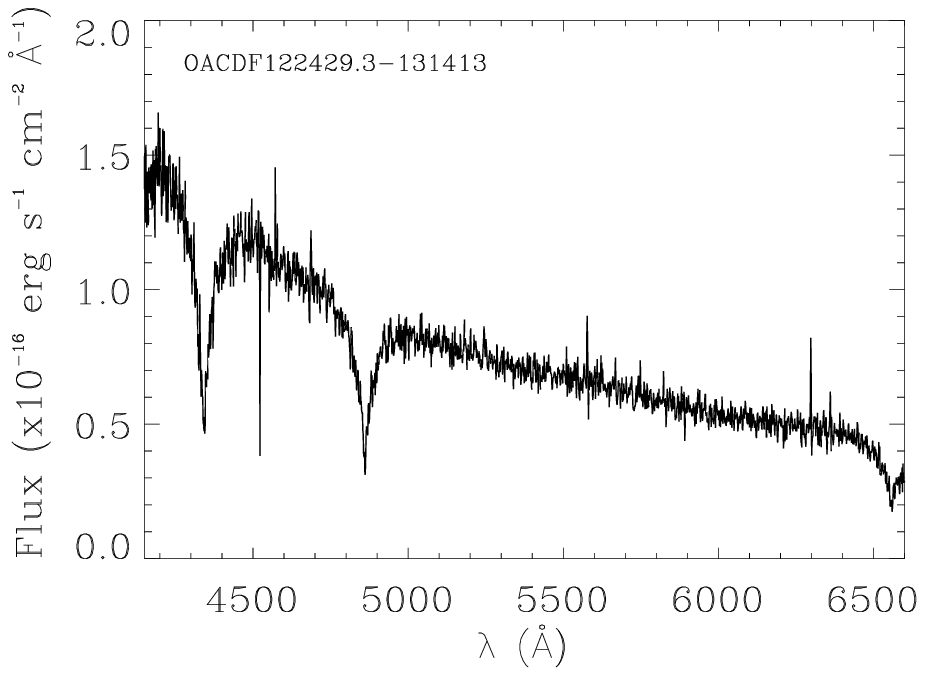}
\includegraphics{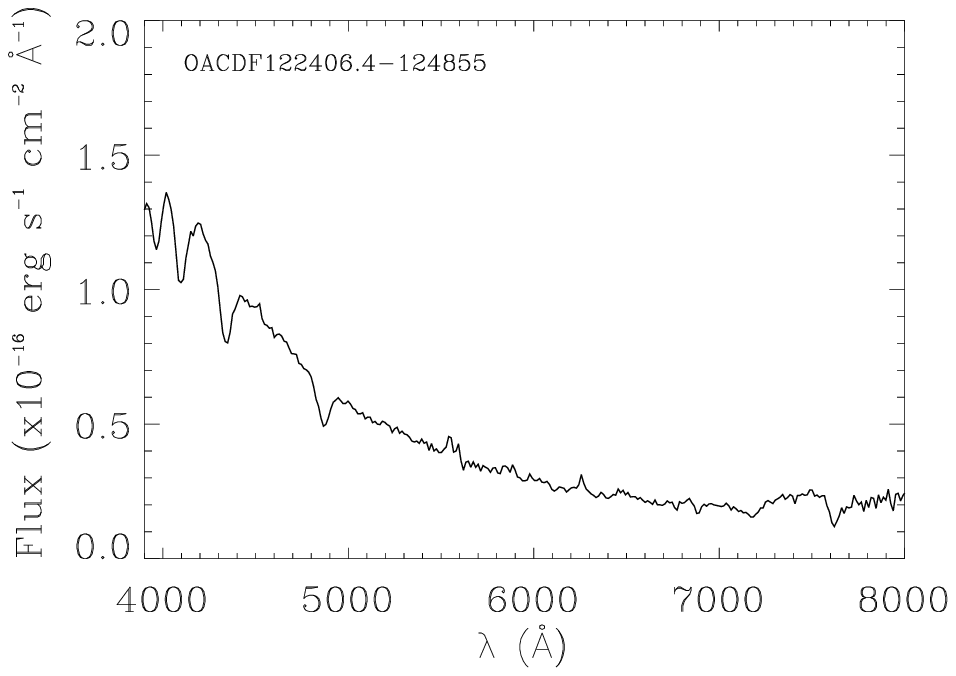}
\caption[]{Flux calibrated spectra of the two confirmed white dwarfs.}
\label{fitswd_1}
\end{figure}

The data were reduced using the standard procedures within the 
ESO-MIDAS~\footnote{ESO Munich Image Data Analysis 
System ({\it http://www.eso.org/sci/data-processing/software/esomidas/}).}
reduction package.
First the images were corrected for bias and flat field;
then the spectra were extracted and wavelength calibrated using arc lamp 
observations; finally, they were flux-calibrated.
From an inspection of the spectra, two objects with the unique presence of 
Balmer lines have been confirmed to be DA white dwarfs.
Their flux calibrated spectra are shown in Fig.~\ref{fitswd_1}.
The third object (which has the up-right position in Fig.~2) has colors 
compatible both with a 8,500~K WD and a A8 main sequence star (e.g. Kenyon \& 
Hartmann 1995). However, its H$\beta$ and H$\gamma$ lines are narrower than 
typical WDs and therefore this star was identified as a A8 star. 
Its spectrum is shown in Catal\'an et al. (2007).

\begin{figure}[h,t]  
\black
\vspace{12.5cm}
\includegraphics{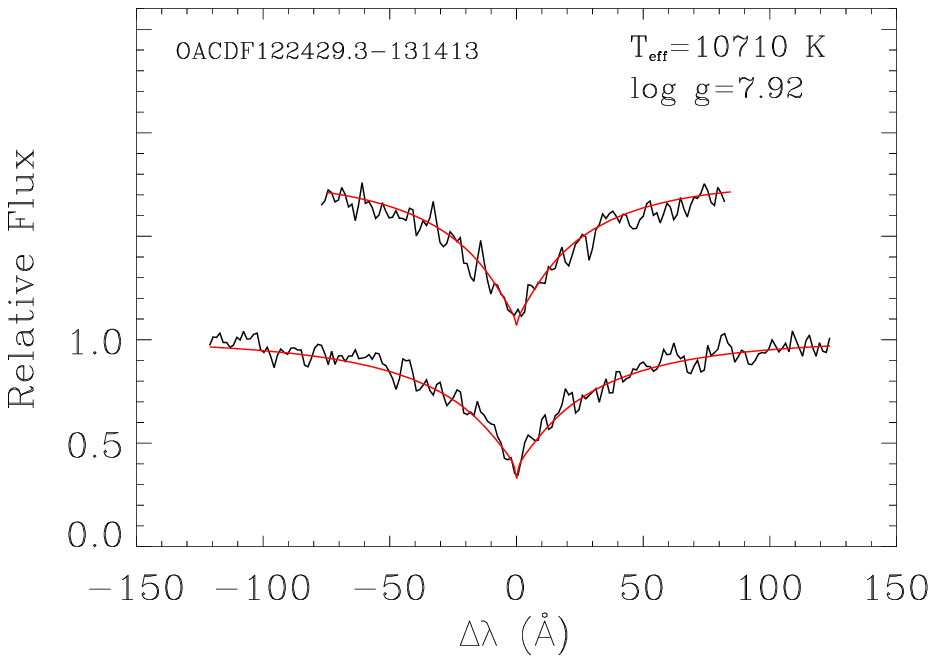}
\includegraphics{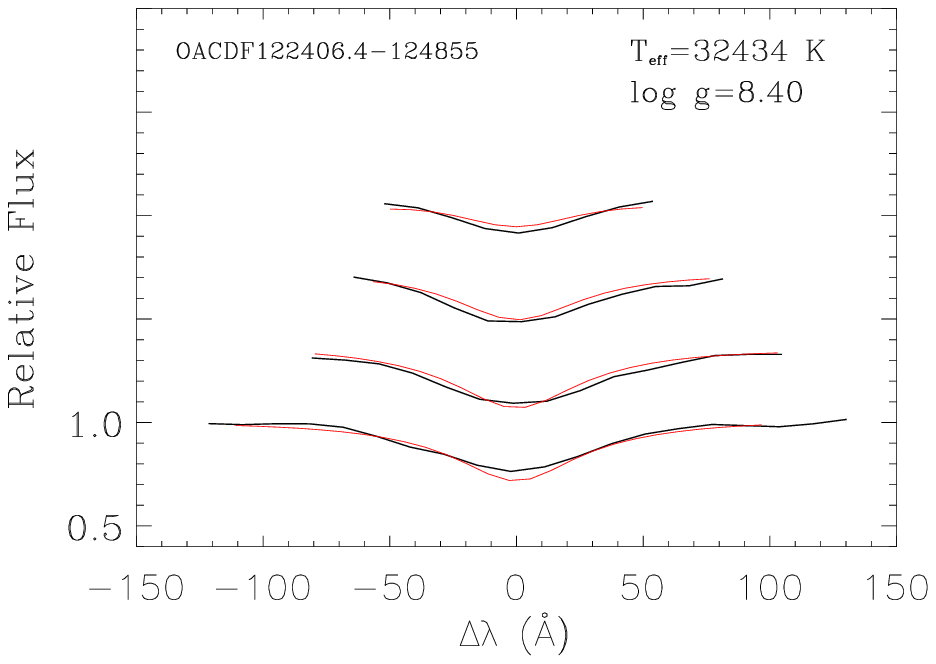}
\caption[]{Model fits (sharp lines) to the individual Balmer line profiles of 
the two DA WDs detected in this survey. Lines range 
from H$\beta$ (bottom) to H$\epsilon$ (top) in the case of 
OACDF122406.4-124855
and
from H$\beta$ (bottom) to H$\gamma$ (top) in the case of OACDF122429.3-131413.
T$_{\rm eff}$ and $\log g$ from the best fit are indicated.}
\label{fitswd_2}
\end{figure}

The effective temperature (\teff) and surface gravity (\logg) of the
two confirmed WDs have been obtained 
following the prescriptions described in Catal\'an et al. 2007.
The method consists mainly in normalizing the spectra to the continuum
and fitting the theoretical models of DA 
white dwarfs by D. Koester (private communication) to the observed Balmer lines
using the package SPECFIT under IRAF~\footnote{Image Reduction and Analysis 
Facility, written and supported by the 
National Optical Astronomy Observatories (NOAO) in Tucson, Arizona 
({\it http://iraf.noao.edu/}).}.
This package is based on $\chi^2$ minimization using the method of 
Levenberg-Marquardt (Press et al. 1992).
The white dwarf models had been previously normalized to the continuum and 
convolved with a Gaussian instrumental profile with the proper FWHM in order to
have the same resolution as the observed spectra. 
In Fig.~\ref{fitswd_2} the fits of the white dwarf models (sharp lines) to the 
observed Balmer lines of the two confirmed white dwarfs are shown. 
In the case of OACDF122406.4-124855 the spectral range is from H$\beta$ to 
H$\epsilon$, while in the case of OACDF122429.3-131413 the spectral coverage 
is narrower, including H$\beta$ and H$\gamma$ only.

\begin{table*}[h,t,b]
\caption{Photometric and stellar parameters of the spectroscopically confirmed 
white dwarfs}
\smallskip
\begin{center}
{\small}
\begin{tabular}{ccccccccccc}
\hline
\hline
\noalign{\smallskip}
\hspace{-2mm}Name & $V$ & \hspace{-1.0mm}$B-V$ & \hspace{-1.5mm}$V-R_C$ &
 \hspace{-0.7mm}$V-I_C$ & \hspace{-1.7mm}$T_{\rm eff}$~(K) &
 \hspace{-2mm}$\log g$~(cgs) & \hspace{-1.8mm}$M_{\star}$~($M_{\odot})$ &
 \hspace{-1.8mm}$t_{cool}$~(Gyr) & \hspace{-2.0mm}$d$~(pc)$^*$\\
\noalign{\smallskip}
\hline
\noalign{\smallskip}
\hspace{-2mm}OACDF122406.4-124855 & \hspace{-1.7mm}19.44$\pm$0.03 &
 \hspace{-2.8mm}--0.10$\pm$0.05 & \hspace{-3mm}--0.02$\pm$0.04 &
 \hspace{-1.4mm}0.61$\pm$0.06 & \hspace{-1.7mm}32,400$\pm$1,300 &
 \hspace{-2mm}8.40$\pm$0.90 & \hspace{-1.8mm}0.88$\pm$0.34 &
 \hspace{-1.8mm}$<$0.13       & \hspace{-2.0mm}620$\pm$$^{610}_{350}$\\
\hspace{-2mm}OACDF122429.3-131413 & \hspace{-1.7mm}19.57$\pm$0.04 &
 \hspace{-1.2mm}0.24$\pm$0.07   & \hspace{-1.2mm}0.04$\pm$0.06 &
 \hspace{-1.4mm}0.03$\pm$0.07 & \hspace{-4.5mm}10,700$\pm$300   &
 \hspace{-2mm}7.92$\pm$0.05 & \hspace{-1.8mm}0.56$\pm$0.02 &
 \hspace{-1.8mm}0.44$\pm$0.04 & \hspace{-2.0mm}350$\pm$20\\
\noalign{\smallskip}
\hline
\hline
\end{tabular}
\end{center}
$^*$
Taking into account galactic extinction the distances are reduced by 20-25 pc.
\end{table*}

Once we have defined \teff~ and \logg~ for each star, we determine its mass 
(M$_{\star}$)
and its cooling time (t$_{cool}$, time elapsed since the planetary nebula
formation) using the cooling sequences of Salaris et al. (2000).
Moreover, comparing the apparent V magnitude with the absolute magnitude
expected from Bergeron's models, we obtain an estimate of the distance.
Our results are shown in Table~2.

The large errors associated with the hotter white dwarf are due to the low 
resolution of the spectrum, implying poor Balmer lines fitting. 
The formal error of 1,300~K in \teff~ is probably underestimated and the 
$B-V$ index would suggest a lower effective temperature near 23,000-24,000~K
(considering a single WD).
Even though the location of this object in Fig.~2 is compatible with
a DB white dwarf, its spectrum does not show any He line, confirming its DA
nature. The $V-R_C$ excess is likely due to a cooler companion, as
confirmed by the high value of $V-I_C$ (see Table~1).


\section{Comparing the WD candidates with synthetic stellar populations}

Taking advantage of the stellar evolution theory \emph{and} galactic structure,
synthetic color-magnitude diagrams (CMDs) can be very useful to disentangle the
stellar counts.
Examples of this kind of
analysis can be found in Bahcall \& Soneira (\cite{BS}), Castellani et
al. (\cite{cast02}), Robin et al. (\cite{robin03}), Girardi et
al. (\cite{girardi}). Here we use the galactic model described in Cignoni et
al. (\cite{cig08}).

Basically, field stars are not randomly distributed along the line of sight,
but they are clumped according to the major galactic 
components (thin disk, thick disk, halo). 
Star count models exploit this point: although the distances are unknown, 
some assumptions can be made on the underlying spatial distributions. 
The following step is to 
convolve the spatial distribution with the underlying stellar populations.
Once the number of synthetic stars for each galactic population at each 
distance from the Sun is computed, masses and ages are extracted populating 
specific initial mass functions (IMFs) and star formation rates (SFRs). 
Next, absolute magnitudes and colors are obtained by interpolating stellar 
tracks, whose metallicity is fixed by the assumed age-metallicity relation. 
Finally, reddening and photometric errors are applied to the synthetic 
photometry.

When a synthetic CMD is ready for a specific observed field
(and scattered as a result of the photometric errors), it is
straightforward to determine CMD regions where the white dwarfs are 
distinguishable from normal stars. In practice, depending on galactic 
latitude and WD colors, a galactic model is used to estimate the probability 
of contamination by: 1) halo turn-off stars; 
2) halo blue Horizontal Branch (HB) stars; 3) massive thin 
disk stars in main sequence (which outnumber the WDs and have similar colors).
Moreover it is possible to estimate the number of expected white dwarfs
and distinguish thin disk white dwarfs from thick disk and halo WDs.

In our model the Galaxy is described by the following ingredients:
\begin{itemize}
\item 
  {\emph{Spatial distributions:}}
The simulated Galaxy includes three main structures, namely the thin disk, 
the thick disk and the halo.
The thin disk and the thick disk density laws are modeled by a double 
exponential (Reid \& Majewski 1993): the main parameters governing these 
profiles are the 
scale length (fixed at 3500 pc for both populations) and the scale height
(1 kpc for the thick disk, 250-300 pc for the thin disk).
The halo follows a power law decay with exponent 3 (see Cignoni et 
al. \cite{cig07}) and an axis ratio of 0.8 (Gould et al. \cite{gould98}). 
A local spatial density of 0.11 stars $\mathrm{pc^{-3}}$ (Reid, Gizis \& 
Hawley 2002) is adopted for the thin disk, whereas thick disk and halo 
normalizations are respectively 1/20 (Robin et al. \cite{robin96}) and 1/850 
relative to the thin disk (Minezaki et al. \cite{minezaki98}, Morrison et al.
\cite{morrison});

\item {\emph{Stellar tracks:}} 
For masses above $0.6\,M_{\odot}$, our code
makes use of the Pisa stellar tracks (Cariulo et al. \cite{car}), while, for
the low mass range, colors are empirically fit to the faint end of nearby
stars (thin disk) and the faint end of the cluster 47-Tucanae (thick
disk). For the halo, the low mass range is treated using the theoretical
calculations by Baraffe et al. (\cite{baraffe}). To predict the CMD location
of WDs we adopt the following ingredients: 1) a WD mass - progenitor mass
relation (Weidemann et al. 2000); 2) WD cooling sequences (Salaris et
al. \cite{salaris00}); 
3) suitable model atmospheres (for
$T_{eff}<4000\,\mathrm{K}$ the color relations are from Saumon \& Jacobson
\cite{saumon}, whereas for higher temperatures the calculations are from
Bergeron, Wesemael \& Beauchamp \cite{bergeron95}). The WD cooling age is
given by the difference between the age of the star and the age of the WD
progenitor at the end of the AGB.
A further point of evolutionary significance concerns the CMD position of the
synthetic HB stars: due to the unknown mechanism of mass loss during the red
giant phase, halo stars which are predicted along the horizontal branch have
been treated by assuming a Gaussian mass distribution centered on a mean mass
$<M_{HB}>$, together with a mass dispersion $\sigma_{M}$ (see e.g. Castellani
et al. \cite{cast05}). Masses and ages are then interpolated using HB stellar
tracks.

\item {\emph{SFRs and chemical compositions:}} The thin disk SFR is assumed
constant and only the old SFR cut-off is allowed to vary between 2 and 6 Gyr
(see e.g. Cignoni et al. \cite{cig06}). The thin disk metallicity is fixed at
$Z=0.02$. SFR and metallicity for halo and thick disk stars are derived from
the comparison between synthetic and observed CMD. 
In the interval $19<V<23$
the CMD data shows a sharp cut-off in color ($0.38<B-V <0.45$). 
This feature can be reproduced 
using a galactic halo with metallicity $Z=0.0008$ (see
also Cignoni et al. \cite{cig07}) and age 11-13 Gyr. Brighter than 19th
magnitude, the observed turn-off moves to the red: although a solution in
terms of age and metallicity cannot prove to be unique, a classical thick
disk with $Z=0.006$ and a star formation activity older than 5 Gyr 
seems appropriate.
 
\item {\emph{IMFs:}} The IMF is a power law ($m^{-\alpha}$) for all
populations: as a range of variation we assume the uncertainty on the exponent
$\alpha$ as evaluated by Kroupa et al. \cite{kroupa} ($\alpha=2.3\pm0.3$).

\end{itemize}

\begin{figure}
  \centering
  \includegraphics[angle=0,width=6cm]{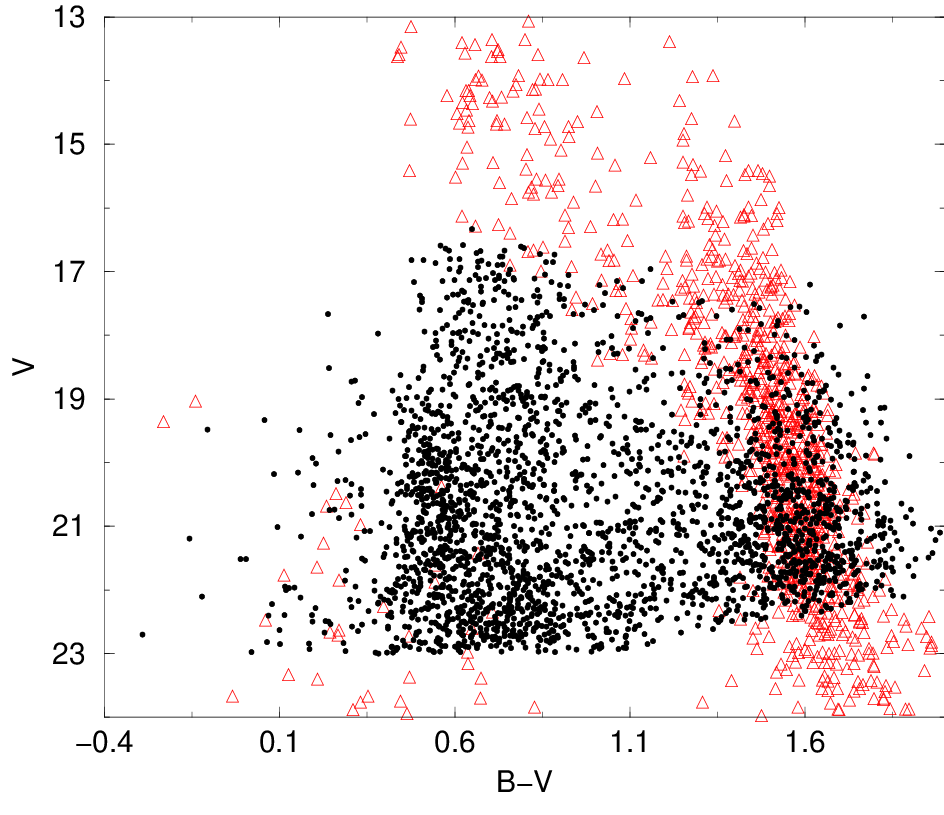}\\
  \includegraphics[angle=0,width=6cm]{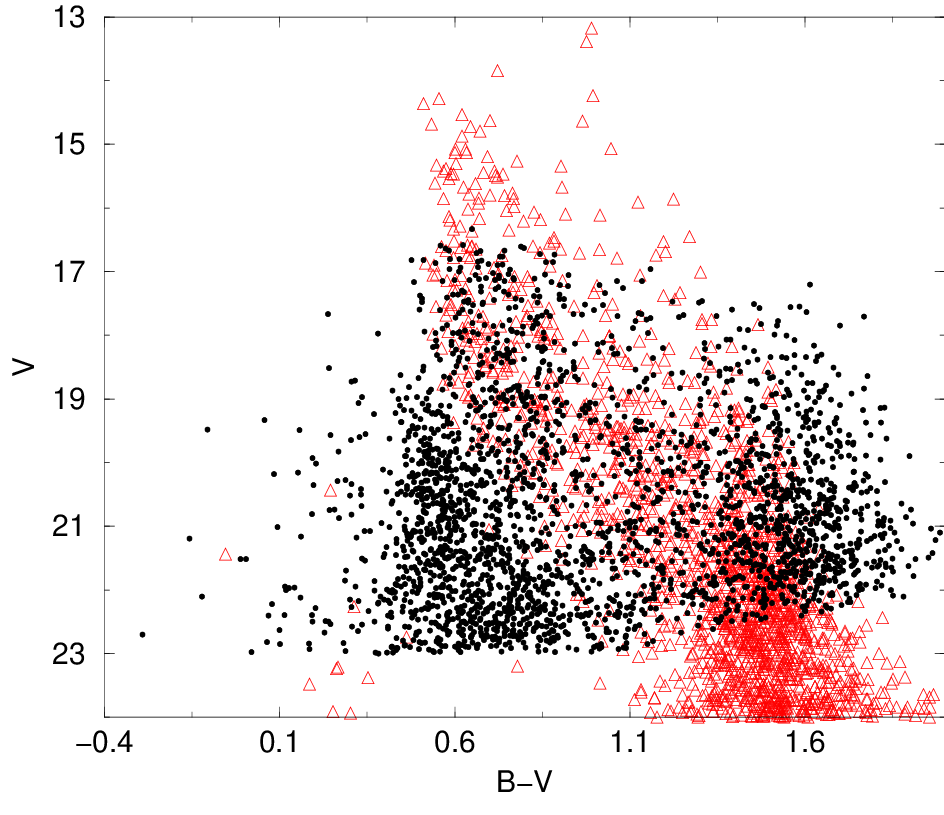}\\
  \includegraphics[angle=0,width=6cm]{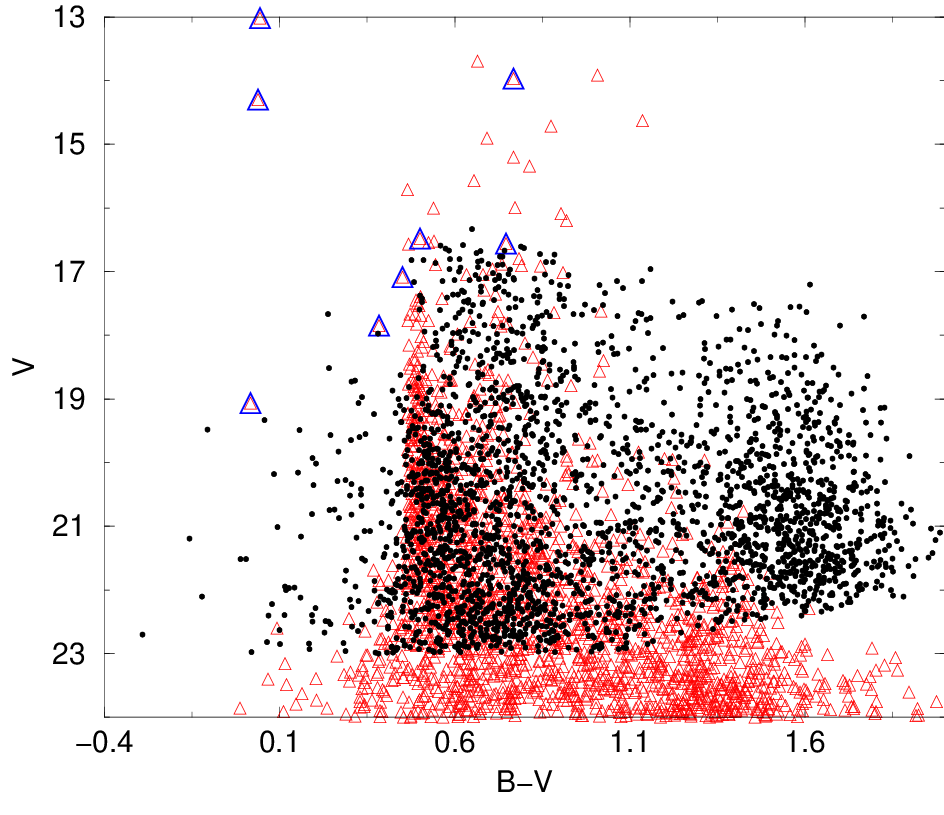}
\caption[]{Color-magnitude diagram of the OACDF point source objects
(black dots) compared with synthetic populations (triangles) of thin disk
(upper panel), thick disk (middle panel) and halo (bottom panel) stars. 
Simulated halo HB stars (emphasized with big triangles) are computed using an
HB mass dispersion $\sigma_{M}=0.005\,M_{\odot}$ and a mean mass
$<M_{HB}>=0.68\,M_{\odot}$.}
\label{cmd_full}
\end{figure}

\begin{figure}[h]  
\vspace{7.8cm}
\includegraphics{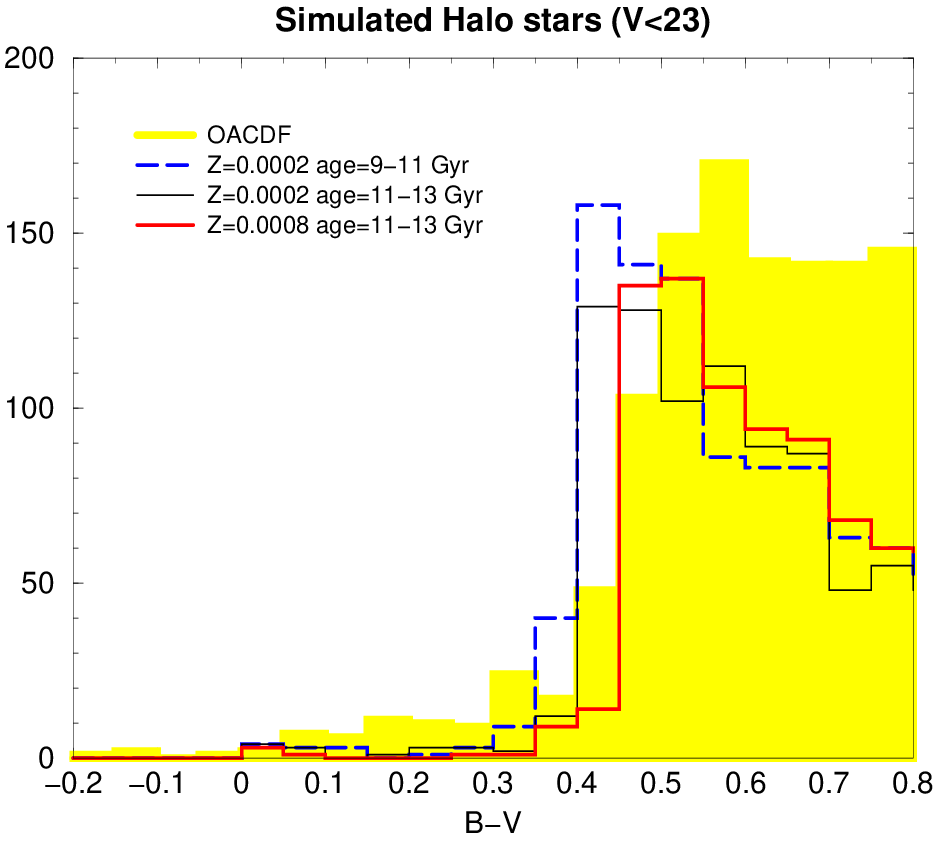}
\caption[]{Comparison between synthetic (only halo stars) and observed color
distribution ($V<23$). Different curves represent different halo models
(HB stars follow the same prescription as in Fig.~5, 
$\sigma_{M}=0.005\,M_{\odot}$ and $<M_{HB}>=0.68\,M_{\odot}$). 
The yellow excess at $B-V$~\lsim~$0.35$ is largely due to the disk white 
dwarfs.}
\label{cmd_halo}
\end{figure}

Figure~\ref{cmd_full}
shows the best synthetic diagram (only single stars,
photometric errors are included) over-plotted onto 
the CMD data. The ingredients are indicated in Tab.~2. 
\begin{table}[t] \centering
\caption[]{Ingredients for the galactic model. The IMFs are from Kroupa et
 al. \cite{kroupa}.}
\begin{tabular}{cccccc}
\hline
\hline
           & Z      & SFR(constant) & spatial parameters (H,L)\\
\hline
DISK       & 0.02   & 0.1-6 Gyr & H=250 pc, L=3500\\
\hline
THICK DISK & 0.006  & 5-10 Gyr  & H=1 Kpc, L=3500\\
\hline
HALO       & 0.0008 & 11-13 Gyr & $\rho \propto R_{Gal}^{\,-3}$\\
\hline
\hline
\end{tabular}
\end{table}
All stars bluer than the halo turn-off are compatible with thin disk white
dwarfs.
However, it is crucial to evaluate any source of contamination.
First, different halo models were considered: although the combination
$Z=0.0008$ and age 11-13 Gyr reproduces well the turn-off region, to isolate a
pure WD sample it is essential to explore also a metal poor halo (which may
dominate sufficiently far out, see e.g. Carollo et al. \cite{carollo}). 
In Figure~\ref{cmd_halo} three halo models are
compared with the observed color distribution (only stars with $V$~$<$~23 are
selected): interestingly, although the turn-off region is sensitive to the
particular model, it is clear that a conservative cut at $B-V\sim 0.35$
significantly reduces the contamination of halo stars.
In particular, the turn-off stars, which make the major contribution, 
never exceed 13 objects, while only few HB stars are present. To evaluate the
maximum HB contamination, the HB mass dispersion $\sigma_{M}$ is varied
between 0.005 and $0.02\,M_{\odot}$,
whereas the mean mass $<M_{HB}>$ is
changed in the range $0.70-0.60\,M_{\odot}$.
The simulations indicate that no
more than 8 HB stars (where this number is obtained with a mean mass of
$0.6\,M_{\odot}$, which is quite extreme) are expected in our field for
$B-V<0.35$ and that they are all brighter than $V$$\sim$19.
It is worth noting that about 25\% of them are ``lost''
because they are brighter than the saturation limit ($V$$\sim$16 for the OACDF).
So the maximum number of visible HB stars is about 6. In our list of OACDF WD
candidates we have 2 stars brighter than $V$=19,
one of them being indeed identified as a HB halo object (see sections 3 and 5
for more details).

As a final remark, we note that a few blue objects may be thick disk WDs: to
test this circumstance, the model thick disk normalization is varied between
5\% and 10\% (which covers most of the current uncertainty). According to the
simulations, the expected number of thick disk WDs with $B-V<0.35$ and $V<23$
is lower than 6-7 objects.

In summary, following our best simulations in Fig.~5, 
besides the thin disk WDs, the CMD region defined by $B-V<0.35$ and 
$16<V<23$ may host a few thick disk WDs ($\sim$3, maximum 6) and about 4
(up to 19 in the worst case) among turn-off and HB halo stars.

In order to provide a more quantitative analysis concerning the WD star counts,
we have compared the observed number of WDs with the model predictions. 
Tab.~3 shows the predicted numbers assuming different IMFs, SFRs
and thin disk scale heights.
As expected, the WD counts increase adopting a shallower IMF, a younger SFR 
and a larger thin disk scale height. 
\begin{table}[t]
\centering
\caption[]{Mean number of predicted thin disk WDs with $B-V<0.35$ and $V<23$,
for different combinations of IMF (exponent $\alpha$) and SFR (constant in
the indicated interval). Results for a thin disk scale height H=250 pc and
H=300 pc are shown.}
\begin{tabular}{ccccc}
&&\bf H=250 pc&\\
\hline
$\alpha$     & N( SFR 0-6 Gyr) & N(SFR 0-4 Gyr) & N(SFR 0-2 Gyr) \\
\hline
1.8          &   26     &         37      &            51\\
\hline
2.3          &   12      &         13     &            23\\
\hline
2.7          &   8       &         11     &            13\\  
\hline
&&&\\
&&\bf H=300 pc&\\
\hline
$\alpha$     & N( SFR 0-6 Gyr) & N(SFR 0-4 Gyr) & N(SFR 0-2 Gyr) \\
\hline
1.8          &   40      &         53      &            77\\
\hline
2.3          &   22      &         23      &            31\\
\hline
2.7          &   13       &      15      &      19\\  
\hline
\end{tabular}
\end{table}
From these simulations, 
if one assumes canonical values for the IMF exponent
($\alpha=2.3$) and the SFR (constant between 0 and 6 Gyr),
we obtain 22 thin disk white dwarfs for a thin disk scale height H=300~pc.
Adding also a few ($\sim$3) thick disk WDs, the total number of white dwarfs 
expected is not far from 32, as derived in section 2.2.
Note that
our list of WD candidates contains at least 4 (more likely 8-10, see Fig.~2) 
WDs in binaries, not included in our simulations.


\section{Discussion}

\begin{figure}[tbh]  
\vspace{8.6cm}
\includegraphics{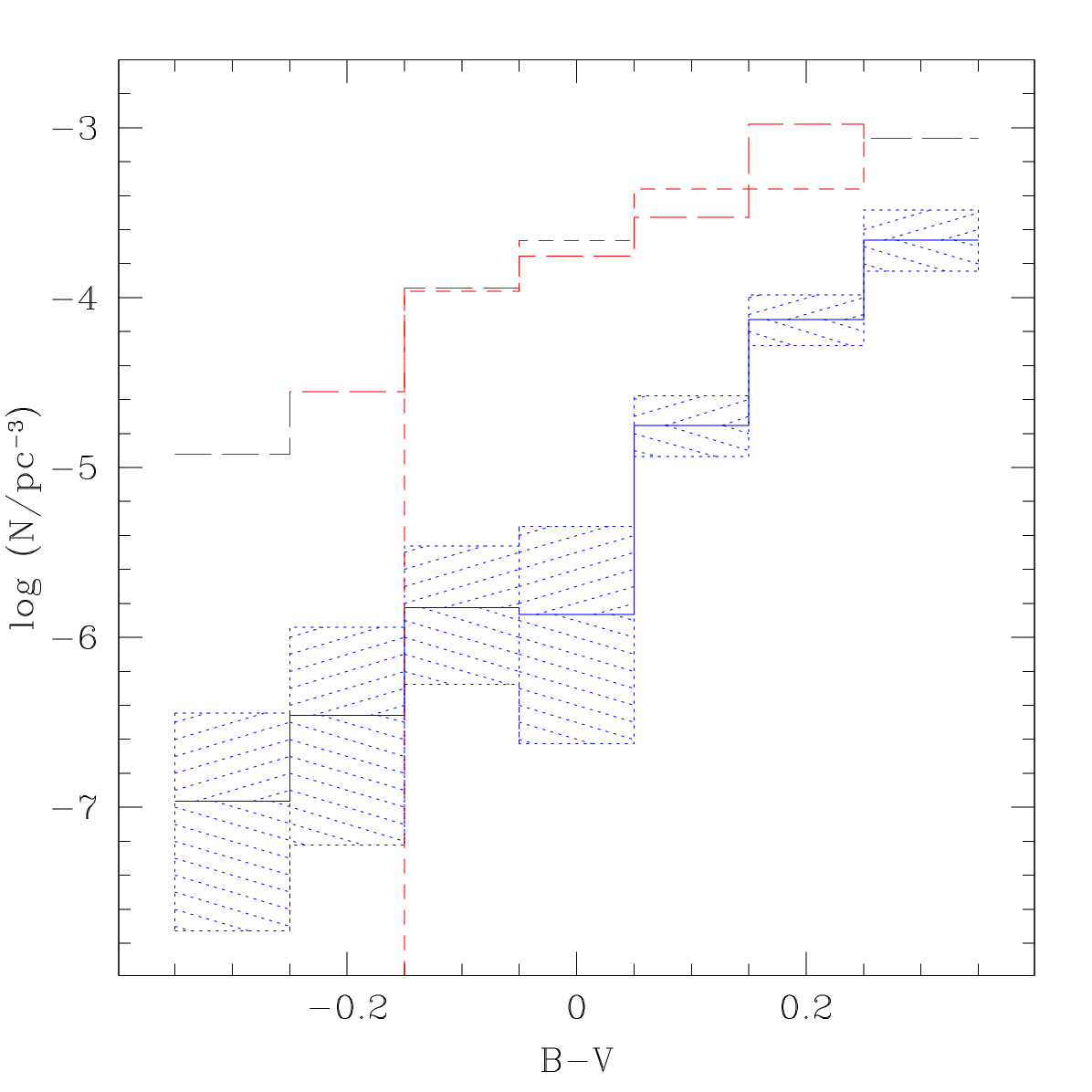}
\caption[]{WD space density at various values of $B-V$.
The continuous line is 
the space density obtained from our data with a V magnitude limit of 23 
(shaded areas represent the errors).
The long and short dashed lines represent, respectively, the WD space density 
expected from Begeron's DA models with \logg =8.0 (Holberg \& Bergeron 2006 
and references therein), taking into account the mean duration of each 
evolutionary phase, and the WD local space density of Holberg et al. (2008)
considering only objects with $B-V<0.35$.
At $0.25<B-V<0.35$ the model density is normalized at the value of 
Holberg et al. (2008); note that their sample does not contain any hot white
dwarf with $B-V<-0.15$.
The great difference in number densities between our and Holberg's data is
due to the fact that we are sampling much more distant regions, in which
the WD space density is much lower (see text for more details).}
\label{isto1}
\end{figure}

\begin{figure}[tbh]  
\vspace{8.6cm}
\includegraphics{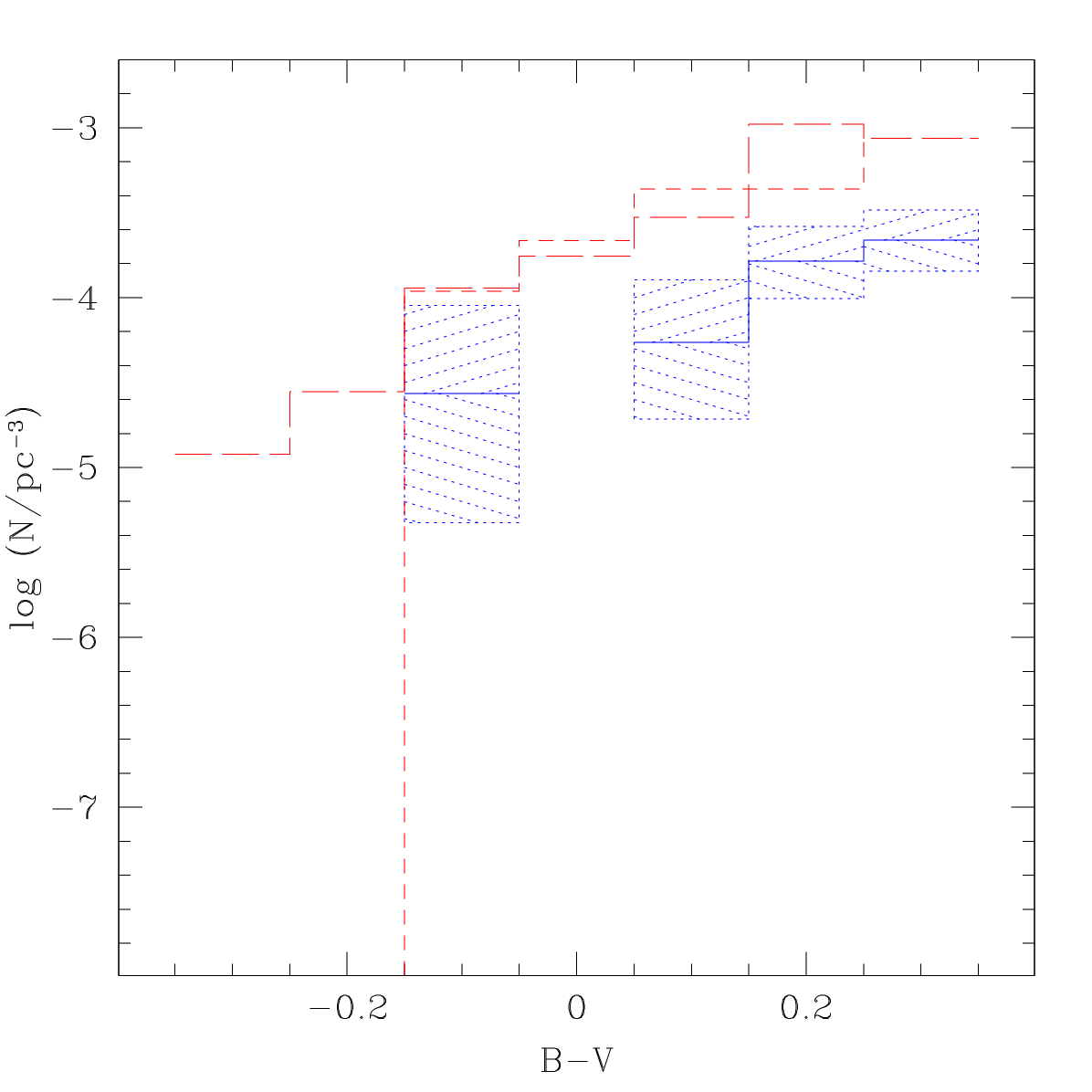}
\caption[]{Same as Fig.~7 but with same volumes sampled in all the color bins.
The continuous (blue) line is the space density within 897~pc from the Sun 
(shaded areas represent again the errors).
In this plot the lower density of bluer objects is related only to
their faster evolution.}
\label{isto2}
\end{figure}

In the previous section we have seen that the number of WD candidates 
that we have found is comparable with the number of disk white dwarfs
predicted by
%
%
synthetic color-magnitude diagrams (including a few thick disk WDs).
In order to confirm the WD nature of these candidates,
spectroscopy was performed for three of them:
one is a non-degenerate star of spectral class A8.
The color of this object ($B-V=0.24$), bluer than the halo turn-off,
and its magnitude ($V=17.7$) suggest that it is a halo star on the horizontal
branch (HB) at 25-30 kpc from us and $\sim$20 kpc from the galactic disk.
This detection is consistent with the synthetic CMDs in Fig.~\ref{cmd_full}
and confirms that stellar contamination is present in our sample.

In order to evaluate also the extra-galactic contamination 
(in particular from QSOs) 
and to compare the number of our WD candidates with previous results, 
we have calculated the space density in different $B-V$ color bins (Fig.~7).
Considering that hotter and more luminous white dwarfs can be seen at larger 
distances, we have calculated the maximum radius at which a white dwarf can be
seen with a magnitude limit $V=23$ for 7 different ranges of $B-V$ 
(absolute magnitudes are calculated using the Bergeron's DA models, 
Holberg \& Bergeron 2006 and references therein, assuming \logg=8.0).
For each bin we have divided the number of WD candidates observed for the 
volume sampled. 
The WD density is then compared with what we expect from Bergeron's DA models,
considering the mean duration of the evolutionary phase corresponding
to that particular color bin, and with the WD local space density obtained by 
Holberg et al. (2008).
Note that what we can compare is only the slope of the density function.  
The number densities are necessarily different because Holberg et al. (2008)
measure the local density (within 13 pc), whereas we are sampling more distant 
regions (see Table 4), up to the WD disk scale height and beyond, where the 
density of white dwarfs is much lower.

Fig.~7 shows that, when we move to hotter white dwarfs, the density 
decreases faster than Bergeron's models and Holberg et al. (2008). 
This is not surprising because hot WDs are observed at larger distances, even 
well beyond the disk (at $B-V=-0.3$ the maximum distance is about 5,700 pc 
at $V=23$), 
where the space density is significantly lower.
When we correct for this effect and recalculate the space densities
considering same volumes in all the colors, the agreement with models
and previous observations is much better (Fig.~8).
This fact suggests that the residual contamination from QSOs in our sample
is relatively small.
Otherwise we would find a greater density in the redder color bin, where the 
frequency of QSOs is higher. 
Using small number statistics (Gehrels 1986), we can estimate a maximum
QSO's contamination of $\approx$50\% in the reddest color bin.

\begin{table*}[h,t]
\caption{Number of WD candidates detected and space densities. The errors are
calculated using small number statistics from Gehrels (1986).}
\smallskip
\begin{center}
{\small}
\begin{tabular}{ccccccc}
\hline
\hline
\noalign{\smallskip}
 & \hspace{-1.5mm}V$_{lim}$=21 & 
   \hspace{-1.5mm}V$_{lim}$=22 & \hspace{-1.5mm}V$_{lim}$=23
& d$<$357pc    & d$<$566pc    & d$<$897pc \\
 & & &
& (z$<$273pc)  & (z$<$434pc)  & (z$<$687pc)\\ 
\noalign{\smallskip}
\hline
\noalign{\smallskip}
               --0.35$<$B--V$<$--0.25 & \hspace{-3.0mm} 0 &  0 &  1 &		      0 &  0 &  0 \\
               --0.25$<$B--V$<$--0.15 & \hspace{-3.0mm} 0 &  1 &  1 &		      0 &  0 &  0 \\
               --0.15$<$B--V$<$--0.05 & \hspace{-3.0mm} 1 &  1 &  2 &		      0 &  0 &  1 \\
\hspace{-1.0mm}--0.05$<$B--V$<$  0.05 & \hspace{-3.0mm} 0 &  1 &  1 &		      0 &  0 &  0 \\
\hspace{0.0mm}   0.05$<$B--V$<$  0.15 & \hspace{-3.0mm} 2 &  4 &  8 &		      0 &  1 &  2 \\
\hspace{0.0mm}   0.15$<$B--V$<$  0.25 & \hspace{-3.0mm} 5 &  6 & \hspace{-1.5mm}11 &  3 &  5 &  6 \\
\hspace{0.0mm}   0.25$<$B--V$<$  0.35 & \hspace{-3.0mm} 5 &  6 &  8 &		      5 &  6 &  8 \\
\noalign{\smallskip}
\hline
\noalign{\smallskip}
TOT & \hspace{-1.5mm}13 & \hspace{-1.5mm}19 & \hspace{-1.5mm}32 & \hspace{-0.5mm}8 & \hspace{-1.5mm}12 & \hspace{-1.5mm}17 \\
\noalign{\smallskip}
\hline
\noalign{\smallskip}
space density (pc$^{-3})$ & & &
& \hspace{-2.0mm}(3.46$\pm$$^{1.71}_{1.20}$)$\times$10$^{-3}$
& \hspace{-2.0mm}(1.31$\pm$$^{0.50}_{0.37}$)$\times$10$^{-3}$
& \hspace{-2.0mm}(4.64$\pm$$^{1.42}_{1.11}$)$\times$10$^{-4}$ \\
\noalign{\smallskip}
\hline
\hline
\end{tabular}
\end{center}
\end{table*}

\begin{figure}[h]  
\vspace{9.4cm}
\includegraphics{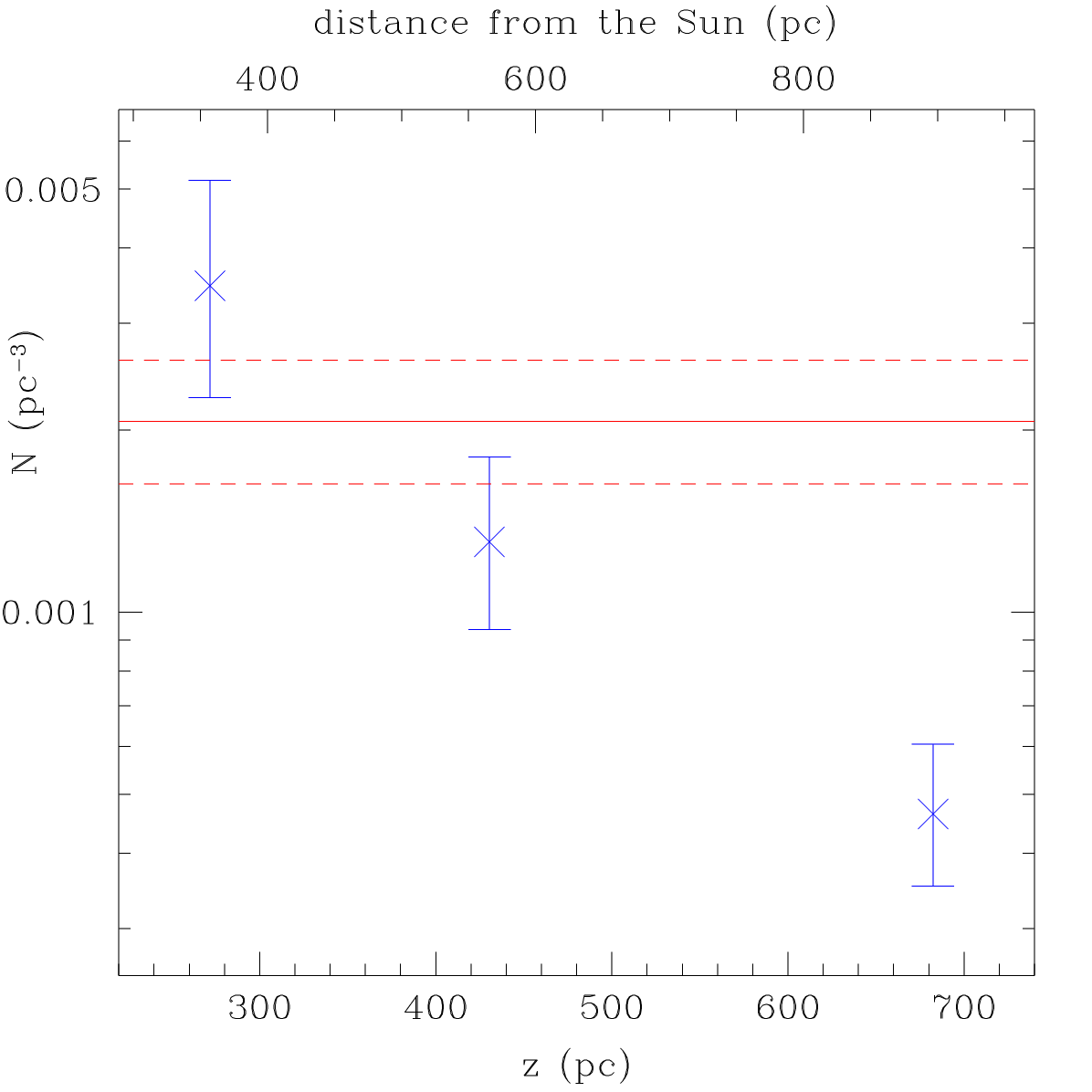}
\caption[]{
White dwarf space density at different distances from the galactic disk (z).
The distribution is compatible with a scale height of about 200 pc.
The horizontal line is the WD local space density (+ errors in dashed lines)
of Holberg et al. (2008) considering only objects with $B-V<0.35$.
}
\label{spacedensity}
\end{figure}

When we sum the space densities over all the color bins, we obtain
the WD space density at various distances from the Sun or from the galactic 
disk (Fig.~9).
The number of WD candidates in various colors and at various distances 
are given in Table~4.
These densities do not include cool WDs with $B-V>0.35$ (or \teff \lsim 7150~K
for DA WDs with \logg =8.0).
In our smallest volume ($d<357$~pc) the WD space density,
summing all the color bins, is 
$(3.5\pm^{1.7}_{1.2})\times10^{-3}$~pc$^{-3}$
(see Table~4), 1.7 times greater than (but within the errors still compatible
to) the local space density of Holberg et al. 2008 
(they obtain $(2.1\pm0.5)\times10^{-3}$~pc$^{-3}$ 
considering only the WDs with $B-V$ between --0.35 and 0.35).
If we consider a $B$ limiting magnitude of 22.5, the number of our WD 
candidates is reduced to 23$\pm$6 in 0.5 square degrees (or 46 per sq. degr.),
which is 1.7$\pm$0.4 times higher than
the sky surface density of 27 degr.$^{-2}$ obtained by Majewski \& Siegel 
(2002) for $B$\,\lsim\,$22.5$ at the north galactic pole.
The effective over-density is reduced from 1.7 to about 1.3 considering the
different galactic latitude of the two fields (the OACDF is at $b\simeq50$).


\vspace{1.5mm}
\noindent
{\it Conclusions:}

\begin{itemize}

\vspace{-2.5mm}
\item
In the OACDF photometric survey we have identified 32 WD candidates. This
number is in agreement (within the errors) with what we obtain from stellar
synthetic populations in the same field.

\item
Our sample may be partially contaminated by blue stars and in particular 
by QSOs in the reddest color bins. 
As described in section 4, the stellar contamination from HB and
turn-off halo stars can be estimated of the order of 10-15\% 
(up to 60\% in very unlikely circumstances).
\hspace{-1mm}
\footnote{
Blue straggler contamination has not been considered. The density
of these stars should be rather small, if not negligible, in low-density 
environments.}
As seen in this section, the QSO contamination is estimated to be 
\lsim50\% in the reddest color bin and almost zero elsewhere, 
giving a contribution of 10-15\% to the total number of white dwarfs. 
Therefore we can consider a total contamination of the order of 30\%.

\item
A selection effect is present for the white dwarfs with red companions.
At $B-V$\,\gsim\,$0.05$ these objects are hardly identified because, 
in the color-color plane, they fall close to or inside the QSO's region 
(Fig.~2).

\item
The WD sky surface density that we find is slightly higher than
($\sim$1.3 times) the value obtained by Majewski \& Siegel (2002).

\item
The WD space density within $\sim$350~pc is 1.7 times greater than 
(but within the errors still compatible to) 
the measurements of Holberg et al. (2008) in the solar 
neighborhood.

\end{itemize}




The results presented in this article represent a small experiment 
in view of similar projects which are planned in the coming years in much wider 
ground-based survey contexts, such as
KIDS/VIKING (see Arnaboldi et al. 2006), STREGA (Marconi et al. 2006) 
and the Alhambra Survey (Moles et al. 2008).


\begin{acknowledgements}

The authors wish to thank  
Jordi Isern, Enrique Garc\'ia-Berro, Detlev Koester, Domitilla de Martino
and Simone Zaggia for discussions and suggestions.
Two visits of S.C. at the Capodimonte Observatory in February and 
September-November 2005 were possible thanks to the MEC grants 
AYA05--08013--C03--01/02 and a research grant associated to the FPU program.
R.S., J.M.A. and A.G. acknowledge financial support from the 
``Regione Campania''.
Several suggestions from an anonymous referee contributed significantly
to improve this article.

\end{acknowledgements}


\end{document}